\documentclass[12pt]{iopart}

\usepackage{iopams}  
\expandafter\let\csname equation*\endcsname\relax
\expandafter\let\csname endequation*\endcsname\relax
\usepackage{amsmath,amssymb}
\usepackage{physics}
\usepackage{bm}
\usepackage{bbm}
\usepackage{hyperref}
\usepackage{color}
\usepackage[dvipsnames]{xcolor}
\usepackage[numbers,sort,compress]{natbib}
\hypersetup{
setpagesize=false,
 bookmarksnumbered=true,%
 bookmarksopen=true,%
 colorlinks=true,%
 linkcolor=Blue,
 citecolor=Blue,
 urlcolor = Blue,
}
\usepackage{graphicx, color}
\newcommand{\eq}[1]{Eq.~(\ref{#1})}
\newcommand{\p}{\partial}
\newcommand{\RM}[1]{\mathrm{#1}}
\begin{document}

\title{Microscopic theory for hyperuniformity in two-dimensional chiral active fluid}

\author{Yuta Kuroda \& Kunimasa Miyazaki}

\address{Department of Physics, Nagoya University, Nagoya 464-8602, Japan}
\ead{kuroda@r.phys.nagoya-u.ac.jp}
\vspace{10pt}
\begin{indented}
\item[]\today
\end{indented}

\begin{abstract}
Some nonequilibrium systems exhibit anomalous suppression of the large-scale density fluctuations, so-called hyperuniformity. 
Recently, hyperuniformity was found numerically in a simple model of chiral active fluids
[\href{https://www.science.org/doi/10.1126/sciadv.aau7423}{Q.-L. Lei {\it et al.}, Sci. Adv.~{\bf{5}}, eaau7423 (2019)}].
We revisit this phenomenon and put forward a microscopic theory to explain it.
An effective fluctuating hydrodynamic equation is derived for a simple particle model of chiral active matter. 
We {show} that the linear analysis of the obtained hydrodynamic equation captures hyperuniformity. 
Our theory yields hyperuniformity characterized by the same exponents as the numerical observation, but the agreement with the numerical data is qualitative. 
We also argue that the hydrodynamic equation for the effective particle representation, in which each rotating trajectory is regarded as an effective particle, has the same form as the macroscopic description of the random organization model with the center of mass conservation.
\end{abstract}

%
%
%
%
%

\section{Introduction}
Active matter {refers to a class of nonequilibrium systems composed} of self-propelled objects, such as {flocks of birds}, bacterial colonies, or self-propelled colloidal particles \cite{Ramaswamy2010Anuual_Review, Marchetti2013RMP, Bechinger2016RMP}.
Each {element in the system} undergoes perpetual motion due to the injection and dissipation of energy, and the system is out of equilibrium at one particle level.
Numerous studies have shown that active matter exhibits many fascinating phenomena forbidden in equilibrium systems.
Examples include
the flocking phase {displaying} long-range polar order in two-dimensional system~\cite{Vicsek1995PRL, TonerPRL1995, Toner1998PRE}; the anomalous dynamics of the topological defect in active nematics~\cite{Sanchez:2012aa, Kawaguchi:2017aa};
spatial-temporal chaotic behavior called active turbulence~\cite{alert2021}; 
self-accumulation without attractive interactions, {known as} motility-induced phase separation (MIPS)~\cite{Taillerur2008PRL, fily2012PRL, Cates2015Annual_Review};
the spatial velocity correlation in the self-propelled particles without alignment interaction~\cite{Henkes2020Nature_Communications, Caprini2020PRL, Caprini2020PRR, Szamel2021EPL, kuroda2023, Caprini2023PRL};
anomalous density or particle number fluctuations often referred to as giant number fluctuations (GNF)~\cite{Ramaswamy2003EPL, Ramaswamy2010Anuual_Review, Marchetti2013RMP}.

GNF has been one of the major concerns in the realm of active matter since  {the} early {days}, and there are a lot of 
experimental \cite{Narayan2007Science,Zhang2010PNAS,Kawaguchi:2017aa, Nishiguchi2017PRE, Iwasawa2021PRR} and numerical observations \cite{Chate2006PRL, Chate2008PRE, Ginelli2012PRL}.
This phenomenon is not observed in thermal equilibrium systems in which the density fluctuations are always short-ranged, except at critical points. 
In nonequilibrium systems, including active matter, the absence of the detailed-balance condition often brings about anomalous increases in density fluctuations at large scales, regardless of the criticality.
GNF is an archetypal example of the large density fluctuations in the system out of equilibrium.
These large density fluctuations are ubiquitous and present even in externally driven simple nonequilibrium fluids~\cite{Dorfman1994, 2006hydrodynamic}, such as 
the Newtonian fluids under the temperature gradient~\cite{Kirkpatric1982,Ronis1982}, constant shear~\cite{Tremblay1981, Nkano2022}, or concentration gradient~\cite{Law1989}.

In contrast, some nonequilibrium systems show the opposite behavior, $i.e.$, 
the density fluctuations are suppressed at large scales even when the particle configuration is disordered.
This is termed hyperuniformity~\cite{TORQUATO}.
Hyperuniformity is characterized by the behavior of the static structure factor 
$S(q)$, where $q$ is the wavenumber.
If this quantity behaves as $S(q)\propto q^{\alpha}$ with a positive exponent $\alpha>0$ in the limit $q\rightarrow 0$, then the system is said to be hyperuniform.
Representative examples of hyperuniform systems {include} the early universe~\cite{Gabrielli2002}, avian photoreceptor cells~\cite{Jiao2014}, periodically driven colloidal suspensions~\cite{Weijs2015,Tjhung_2016}, and jamming systems~\cite{Donev2005,Ikeda2015,Matsuyama:2021}.
Recently, it has been reported that the two-dimensional chiral active fluids, {in which} the trajectory of each particle violates the right-left symmetry, also exhibit hyperuniformity~\cite{Li2019,Lei2019_pnas,Liu2022,Huang2021,Zhang2022}. 
Chiral active matter is abundant in nature~\cite{Lowen2016,Ma2017,Liebchen2017,Banerjee2017, Liebchen_2022}: most self-propelling objects, such as bacteria~\cite{DiLuzio2005, Leonardo2011}, algae~\cite{Vincent2012}, and Janus particles with an asymmetric coating~\cite{Mano2017}, cannot perform completely symmetric motions, and their motion has chirality. 
Hyperuniformity in the two-dimensional chiral active fluids was first observed in a numerical study \cite{Li2019}, where the athermal chiral active particle model was employed. 
This system undergoes the nonequilibrium phase transition from the absorbing state to the active state as the orbital radius or number density increases. 
In the latter state, the system is hyperuniform characterized by $S(q)\propto q^{2}$ (or $\alpha=2$) for the small $q$'s region.
Interestingly, this hyperuniformity  {occurs} far above the critical point, although most systems that undergo the absorbing phase transition, such as random organization models, exhibit hyperuniformity only in the vicinity of the critical point~\cite{Weijs2015,Tjhung_2016}.
The other model that exhibits hyperuniformity far above the critical point is the random organization model with the center of mass conservation (COMC) \cite{Hexner2017, Lei2019_pnas}.
The authors in Ref.~\cite{Li2019} also conjectured that the two-dimensional chiral active particles can be mapped into the COMC dynamics \cite{Hexner2017}, inspired by the identical hyperuniform exponents, $\alpha =2$. 
Active spinner models are also found to show hyperuniformity \cite{Lei2019_pnas,Liu2022}. 
Experimentally, hyperuniformity induced by chirality has been observed in marine algae~\cite{Huang2021} and pear-shaped Quincke rollers~\cite{Zhang2022}.
Despite intensive studies, however, a microscopic theory that explains the chirality-induced hyperuniformity is still absent.

In this paper, we derive a fluctuating hydrodynamic description for the chiral active fluids by a bottom-up approach and {show} that linear analysis of the obtained equations yields hyperuniformity characterized by the same exponent as the numerical observation in Ref.~\cite{Li2019}. 
Our hydrodynamic theory leads to $S(q)$ in the general case where both the torque and persistence time are finite. 
The ``authentic" hyperuniformity, in which $S(q)$ exactly vanishes in the limit $q\rightarrow 0$, can be found only at the infinite persistence time.
In case the persistence time is finite, hyperuniform structure is destroyed.  
However, the suppression of the density fluctuations can be seen over a wide range of wavenumber if the persistence time is sufficiently large compared with the time scale of chiral motion.
We also perform the numerical simulation for a particle model and compare it with our theoretical results.
Although our theory yields the same exponents as the numerical observation, the pre-factor does not agree with the numerical data. 
Our theory predicts that the pre-factor is given solely by the chirality of a particle, but numerical results show that it also depends on the number density.
We also consider the conjecture in Ref.~\cite{Li2019} that the chirality-induced hyperuniformity should be understood in a similar way to the COMC dynamics~\cite{Hexner2017}. 
This conjecture was based on the idea that each rotating trajectory of a chiral particle is regarded as an effective particle.
In the random organization model with COMC dynamics, hyperuniformity {arises} from the noise term with a square gradient which possesses an additional first-order spatial derivative to the standard conserved noise~\cite{Hexner2017}.
Here we propose one of the routes to derive the square-gradient noise for the chiral active particles in the effective particle representation and demonstrate that the conjecture in Ref.~\cite{Li2019} is justified at least in cases where the persistence time is finite. 
 
This paper is organized as follows.
We derive the fluctuating hydrodynamic equations for a chiral active particle model in Sec.~\ref{FH}. 
In Sec.~\ref{sq}, we demonstrate that the linear analysis of the obtained hydrodynamic equations yields hyperuniformity with $\alpha =2$ in the infinite persistence time limit.
To see the validity of the theory, we compare the numerical simulation to our theory in Sec.~\ref{sim}.
We also consider the hydrodynamic description in the effective particle representation in Sec.~\ref{eff}.
Finally, we devote Sec.~\ref{conc} to the conclusion and remarks. 

\section{Derivation of effective fluctuating hydrodynamic equations}
\label{FH}

Our starting point is the two-dimensional chiral active Brownian particles (cABP)~\cite{Ma2017, Li2019}. 
The dynamics of the particle $j\in\{1,2,...,N\}$ is described by the following Langevin equations:
\begin{align}
&\dv{\bm r_j(t)}{t} = \mu\bm F_j(t) + v_0\bm e(\phi_j(t)),  \label{1}\\
&\dv{\phi_j(t)}{t} = \Omega + \sqrt{2D}\eta_j(t) , \label{2}
\end{align}
where $\mu, v_0, \Omega$, and $D$ denote the mobility, self-propelling speed, constant torque, and rotational diffusion constant, respectively.  $N$ stands for the total number of particles.
The orientation of a particle is represented by the unit vector $\bm e(\phi)=(\cos\phi,\sin\phi)$.
 $\eta_j(t)$ is a Gaussian white noise satisfying $\expval{ \eta_j(t)} =0$ and $\expval{\eta_j(t)\eta_k(t')}=\delta_{j,k}\delta(t-t')$. 
 $\bm F_j(t) = -\sum_{k=1}^{N}\nabla_jU(\abs{\bm r_j - \bm r_k})$ represents the  {repulsive} interaction force between the particle $j$ and other particles. Here, we assume the pair potential satisfies $\eval{\nabla U(r)}_{ r=0}=\bm 0$ for simplicity.
If $\bm F_j(t)$ and $D$ are absent, each particle performs the circular motion with the radius $R=v_0/\Omega$.
 If $\bm F_j(t)$ is absent but $D$ is finite, the particles undergo a diffusive motion with the diffusion constant $D_\Omega = D_\RM{act}/[1+(\Omega/D)^2]$, where $D_\RM{act} = v_0^2/(2D)$ (see \ref{apeA}).
 In the limit $\Omega \rightarrow 0$,  Eqs.~(\ref{1}) and (\ref{2}) are reduced to the equations for the standard active Brownian particles (ABP)~\cite{fily2012PRL,Bechinger2016RMP} in which each particle performs the ballistic motion for the persistence time $1/D$.
In the limit $D\rightarrow \infty$, the system becomes the equilibrium system described by the overdamped Langevin equation with the temperature $T_\RM{act}=v_0^2/(2D\mu)$ because in this limit, the active force $v_0\bm e(\phi_j)$ becomes a Gaussian white noise, and the fluctuation-dissipation relation is recovered (see \ref{apeA}).
Note that the original model in Refs.~\cite{Li2019,Ma2017} includes the thermal noise in \eq{1}, but here we omit this term for simplicity. 
 {The fluid states of cABP exhibit several different states depending on the parameters~\cite{Li2019, Ma2022}.
When the rotational diffusion constant $D$ is small, and the density is high enough, the system undergoes clustering.
When $D=0$, the systems fall into absorbing states at very small $R$ or density.
Namely, cABP has three distinct states, clustering, absorbing, and disordered homogeneous state, at $D=0$. 
In the disordered homogeneous state,
hypueruniformity characterized by $S(q)\propto q^2$ occurs.
Here, we focus only on the disordered homogeneous state.
}

 {
The goal of this section is to construct an effective fluctuating hydrodynamic description for the homogeneous fluid states from Eqs.~(\ref{1}) and (\ref{2}).
Several methods to drive effective hydrodynamic equations have been proposed for active fluids. 
One of the standard methods is to derive a set of closed equations for the averaged hydrodynamic fields with an appropriate closure procedure of the hierarchical equations~\cite{Bialk2013, Speck2014PRL, Speck2015, Speck2021pre}.
This method has been applied to chiral active fluids to investigate the instabilities~\cite{Li2019, Bickmann2022, Ma2022, Sansa2022, Kreienkamp_2022}. However, since our interest is fluctuations of hydrodynamic fields in the homogeneous states, we need to construct the Langevin equations for the hydrodynamic fields.
A naive way to deal with the fluctuations is to add Gaussian white noise into the averaged equations.
However, for active fluids, there is no guarantee that such noise terms are correct.
To construct the equations with noise terms, we use Dean's method~\cite{Dean_1996, Nakamura_2009}, which is a standard way of changing microscopic stochastic variables such as particle positions to fluctuating hydrodynamic fields such as the density field.
}

The hydrodynamic fields of this system are the density and polarization fields defined by
 \begin{align}
\rho(\bm r,t) &= \sum_{j=1}^{N}\delta(\bm r-\bm r_j(t)), \label{3}\\
\bm p(\bm r,t) &= \sum_{j=1}^{N} \bm e(\phi_j(t))\delta(\bm r-\bm r_j(t)), \label{4}
\end{align} 
respectively.
First, we construct the equation for density field \eq{3}. 
From the time derivative of \eq{3}, we obtain the continuum equation as
\begin{equation}
\p_t{\rho(\bm r,t)} = -\nabla\cdot \bm J(\bm r,t), \label{5}
\end{equation}
where the current $\bm J(\bm r,t)$ is given by
\begin{align}
\bm J(\bm r,t) &:= \sum_{j=1}^{N}\dot{ \bm r} _j(t) \delta(\bm r-\bm r_j(t)) \notag \\
&= \mu \sum_{j=1}^{N}\bm F_j(t) \delta(\bm r-\bm r_j(t))  + v_0\bm p(\bm r,t). \label{6}
\end{align}
The interaction term on the right-hand side of \eq{6} can be expressed as
\begin{align}
\sum_{j=1}^{N}\bm F_j(t)\delta(\bm r- \bm r_j(t)) 
= - \rho(\bm r,t)\nabla\fdv{\mathcal F[\rho(\cdot,t)]}{\rho(\bm r,t)},\label{7}
\end{align}
where
\begin{equation}
\mathcal F[\rho(\cdot,t)] := \frac{1}{2}\int_{V}\dd[2]\bm r\int_{V}\dd[2]\bm r'\ \rho(\bm r,t)\rho(\bm r',t)U(\abs{\bm r-\bm r'}). \label{free_ene}
\end{equation}
Next, we derive the equation for the polarization \eq{4}.
The time derivative of \eq{4} leads to
\begin{equation}
\p_t{\bm p(\bm r,t)} = - \nabla\cdot \mathsf M^\RM{e}(\bm r,t)  + \sum_{j=1}^{N}\dv{\bm e(\phi_j(t))}{t} \delta(\bm r -\bm r_j(t)), \label{11}
\end{equation}
where we defined the tensor $\mathsf M^\RM{e}(\bm r,t)$ as
\begin{equation}
\mathsf M^\RM{e}(\bm r,t):=  \sum_{j=1}^{N}\dot{\bm r}_j(t)\bm e(\phi_j(t)) \delta(\bm r -\bm r_j(t)). \label{12}
\end{equation}
Following Ref.~\cite{Nakamura_2009}, to express \eq{12} in terms of the hydrodynamic fields, we assume 
\begin{equation}
\delta(\bm r_j(t) - \bm r_k(t)) = \delta(\bm r_j(t) - \bm r_k(t))\delta_{j,k}. \label{13}
\end{equation}
This relation is correct if the particles $j$ and $k$ never occupy the same position at the same time, such as when the pairwise potential $U( r)$ is repulsive interactions \cite{Nakamura_2009}.
Using the assumption \eq{13}, \eq{12} becomes an advection term:
\begin{align}
\mathsf M^\RM{e}(\bm r,t) 
&= \frac{1}{\rho(\bm r,t)}\sum_{j=1}^{N}\rho(\bm r,t)\dot{\bm r}_j(t)\bm e(\phi_j(t)) \delta(\bm r -\bm r_j(t)) \notag \\
&= \frac{1}{\rho(\bm r,t)}\sum_{j,k=1}^{N}\dot{\bm r}_j(t)\bm e(\phi_k(t)) \delta(\bm r -\bm r_j(t)) \delta(\bm r -\bm r_k(t))\notag \\
&= \frac{\bm J(\bm r,t)\bm p(\bm r,t)}{\rho(\bm r,t)}. \label{14}
\end{align}
From \eq{2} and the It\^o formula \cite{gardiner}, the time derivative of the unit vector $\bm e(\phi_j(t))$ on the right-hand side of \eq{11} is written as
\begin{align}
\dv{\bm e(\phi_j(t))}{t}
= - D\bm e(\phi_j(t)) + \Omega  \mqty(-\sin\phi_j(t) \\ \cos\phi_j(t) ) 
 +\sqrt{2D}  \mqty(-\sin\phi_j(t) \\ \cos\phi_j(t) ) \bullet \eta_j(t) &. \label{15}
\end{align}
Here, the symbol $\bullet$ denotes the It\^o product.  
By substituting Eqs.(\ref{14}) and (\ref{15}) into \eq{11}, we have
\begin{align}
\p_t\bm p(\bm r,t) = -\nabla\cdot\qty(\frac{\bm J(\bm r,t)\bm p(\bm r,t)}{\rho(\bm r,t)})
- D\bm p(\bm r,t) 
+ \bm \Omega \times \bm p(\bm r,t) +\bm \Lambda(\bm r,t)& . \label{16}
\end{align}
For convenience, we introduced the three-dimensional vector $\bm \Omega=(0,0,\Omega)$ to express as the cross product. The noise term $\bm \Lambda(\bm r,t)$ is defined as
\begin{equation}
\bm \Lambda(\bm r,t):=\sqrt{2D}\sum_{j=1}^{N}\mqty(-\sin\phi_j(t) \\ \cos\phi_j(t) ) \bullet \eta_j(t)  \delta(\bm r -\bm r_j(t)). \label{17}
\end{equation}
Note that if we rewrite \eq{16} in the Stratonovich representation, then the dissipation term $-D\bm p(\bm r,t)$ vanishes, and the mean value of the noise term $\bm \Lambda (\bm r,t)$ is non-zero. 
Since \eq{17} depends on the microscopic variables, \eq{16} is still not a closed equation for the hydrodynamic fields. However, in the large $N$ limit, \eq{17} can be rewritten as the following simple form:
\begin{equation}
\bm \Lambda(\bm r,t) = \sqrt{D\rho(\bm r,t)}\bm \Upsilon(\bm r,t), \label{18}
\end{equation}
where $\bm \Upsilon(\bm r,t)$ is a Gaussian white noise satisfying $\expval{\bm \Upsilon(\bm r,t)} = \bm 0$ and
\begin{equation}
\expval{ \Upsilon_\alpha(\bm r,t) \Upsilon_\beta(\bm r',t')} = \delta_{\alpha,\beta}\delta(\bm r-\bm r')\delta(t-t').
\end{equation}
Here Greek indices represent Cartesian coordinates. 
The derivation of \eq{18} is described in \ref{apeB}.
{Combining} the above results, Eqs.~(\ref{5})-(\ref{7}), and (\ref{16}), we reach the set of closed Langevin equations for the density field and polarization.
Note that the obtained equations are different from the Dean--Kawasaki equation~\cite{Dean_1996, Kawasaki1994}. 
In particular, the functional $\mathcal F[\rho]$, \eq{free_ene}, does not contain the diffusive part or so-called ideal part.
The connection to the Dean--Kawasaki equation~\cite{Dean_1996, Kawasaki1994} in the equilibrium limit is described in~\ref{Eqlim}.

Up to now, the manipulations are almost rigorous and there are no assumptions except for \eq{13}. 
Thus the obtained equations are valid not only for the macroscopic length scale but still possess microscopic information. 
Henceforth, we focus only  {on} the macroscopic length scales, or in the hydrodynamic limit.
We expand $\mathcal F[\rho]$ around the mean density $\rho=N/V$ and assume that \eq{7} can be approximated by linear order of the density fluctuation $\delta\rho(\bm r,t) = \rho(\bm r,t) - \rho$ in the hydrodynamic limit:
\begin{equation}
 \rho(\bm r,t)\nabla\fdv{\mathcal F[\rho(\cdot,t)]}{\rho(\bm r,t)} \simeq \frac{1}{\rho\chi}\nabla \delta \rho(\bm r,t), \label{9}
\end{equation}
where ${\chi}$ is a coefficient that corresponds to the compressibility in the hydrodynamics.   
We address that the linearization in \eq{9} is a crude approximation: the left-hand side of \eq{9} can contain the higher gradient terms and coupling between the density field and polarization. 
Derivation of the hydrodynamic equations containing these non-linear terms using systematic coarse-graining procedures is involved and beyond the scope of the present work.
Although our treatment is crude, it turns out to capture the essence of hyperuniformity in this system, at least qualitatively, as we will see below. 
Plugging Eqs.~(\ref{7}) and (\ref{9}) {into} \eq{5}, we obtain
\begin{equation}
\p_t{\rho(\bm r,t)} = \frac{\mu}{\rho\chi}\nabla^2\rho(\bm r,t)  - v_0\nabla\cdot \bm p(\bm r,t) .\label{10}
\end{equation}

Now, we focus on the fluctuation around the stationary state $(\rho(\bm r,t),\bm p(\bm r,t))=(\rho,\bm 0)$. 
Linearizing Eqs.~(\ref{10}) and (\ref{16}) in terms of the density fluctuation $\delta \rho(\bm r,t)$ and polarization fluctuation $\delta \bm p(\bm r,t) = \bm p(\bm r,0)-\bm 0$, we have the linearized equations:
\begin{align}
&\p_t \delta\rho(\bm r,t) = b \nabla^2 \delta\rho(\bm r,t)  - v_0 \nabla\cdot \delta\bm p(\bm r,t) , \label{20}  \\
&\p_{t} \delta\bm p(\bm r,t) = - D\delta\bm p(\bm r,t)  + \bm \Omega \times \delta\bm p(\bm r,t)  + \sqrt{D\rho}\bm \Upsilon(\bm r,t),\label{21}
\end{align}
where we defined $b: = \mu/(\rho\chi)$.
\eq{21} is a closed equation for polarization. 
Therefore, we can regard $\bm \xi(\bm r,t) := v_0 \delta \bm p(\bm r,t)$ in \eq{20} as a colored noise.
This noise has zero mean and the following correlation:
\begin{equation}
\expval{\bm \xi(\bm r,t) \bm \xi^\RM{T}(\bm r',t') } = \frac{v_0^2 \rho}{2} \mqty(\cos[ \Omega (t-t')] & -\sin [ \Omega (t-t')] \\ \sin [ \Omega (t-t')]  & \cos [ \Omega (t-t')] ) e^{-D\abs*{t-t'}} \delta(\bm r- \bm r'),
\end{equation}
where the symbol $\RM{T}$ stands for the transpose operation. 
In the limit $D\rightarrow \infty$ while keeping $T_\RM{act} = v_0^2/(2D\mu)$ constant, the fluctuation dissipation relation is recovered as $\expval*{\bm \xi(\bm r,t) \bm \xi^\RM{T}(\bm r',t') } = 2 \rho \mu T_\RM{act} \delta(\bm r-\bm r')\delta(t-t')\mathbbm{1}$. 

\section{Density correlation function}
\label{sq}
From the linearized fluctuating hydrodynamic equations, Eqs.~(\ref{20}) and (\ref{21}), we can calculate the static structure factor defined by 
\begin{equation}
S( q) = \frac{1}{N}\expval{\delta\tilde{\rho}(\bm q, 0)\delta \tilde{\rho}^*(\bm q, 0)},
\end{equation}
where the variable with tilde $\tilde X(\bm q,t)$ denotes the Fourier transform of $X(\bm r,t)$ with respect to $\bm r$, and $*$ stands for the complex conjugate. 
In this section, we show that in the limit of $D\rightarrow 0$, $S(q)$ is proportional to $q^{2}$ for the small wavenumber. 

In the Fourier space, Eqs.~(\ref{20}) and (\ref{21}) can be written as 
\begin{equation}
\mathsf G^{-1}(\bm q,\omega) \hat{\bm \Phi}(\bm q,\omega) = \hat{\bm \Xi}(\bm q,\omega) ,
\label{23}
\end{equation}
where $\hat{\bm \Phi}(\bm q,\omega)$ and $\hat{\bm \Xi}(\bm q,\omega)$ are the spatiotemporal Fourier transform of
\begin{align}
\bm \Phi (\bm r,t) &:= \mqty(\delta\rho(\bm r,t) & \delta p_x(\bm r,t)& \delta p_y(\bm r,t) )^\RM{T},\\ 
\bm \Xi (\bm r,t) &:=\sqrt{D\rho} \mqty(0 & \Upsilon_x(\bm r,t)& \Upsilon_y(\bm r,t) )^\RM{T},
\label{25}
\end{align}
respectively,
and
\begin{equation}
\mathsf G^{-1}(\bm q,\omega)  := 
\mqty(
- i\omega + bq^2 & iv_0 q_x & iv_0 q_y \\
0 & -i\omega + D& \Omega \\
0 & -\Omega & -i\omega + D
).
\end{equation}
From \eq{23}, we can calculate the dynamical correlation functions in $(\bm q,\omega)$-space given by
\begin{equation}
S^\Phi_{a,b}(\bm q,\omega) = \frac{1}{\rho}\int_{V}\dd[2]\bm r \int_{-\infty}^{\infty}\dd t\ \expval{\Phi_a(\bm r,t)\Phi_b(\bm 0,0)}e^{-i(\bm q\cdot \bm r - \omega t)}. \label{correlation}
\end{equation}
In the stationary states, \eq{correlation} satisfies 
\begin{equation}
    \frac{1}{N}\expval{\hat\Phi_a(\bm q,\omega)\hat\Phi^*_b(\bm q,\omega')} = 2\pi S^\Phi_{a,b}(\bm q,\omega)\delta(\omega-\omega'). \label{W-K}
\end{equation}
Here we focus on only the density correlation function or dynamical structure factor $S(\bm q,\omega)=S^\Phi_{1,1}(\bm q,\omega)$. 
Using \eq{W-K}, $S(\bm q,\omega)$ can be obtained as
\begin{align}
S(\bm q,\omega ) &= D\qty[ \abs{G_{1,2}(\bm q,\omega) }^2 + \abs{G_{1,3}(\bm q,\omega) }^2] \notag \\
&= \frac{v_0^2 D q^2\qty (\omega^2 + D^2 + \Omega^2)}{ \qty(\omega^2 + b^2q^4)\qty[ \qty{ \omega^2 - (D^2 + \Omega^2)}^2 + 4D^2\omega^2]}. \label{28}
\end{align}
By integrating \eq{28} over $\omega$, we obtain $S(q)$ as
\begin{align}
S(q) &=  \frac{1}{2\pi}\int_{-\infty}^{\infty}\dd\omega\ S(\bm q,\omega) =\frac{v_0^2}{2b} \frac{D+ bq^2}{\Omega^2 + \qty(D+bq^2 )^2} \label{29} . 
\end{align}
The behavior of \eq{29} changes depending  on whether $\Omega > D$ or $\Omega <D$.
First, we consider an extreme case, $D\rightarrow 0$ with finite $\Omega$, which corresponds to the noiseless limit studied in Ref.~\cite{Li2019}.
In this case, \eq{29} becomes 
\begin{equation}
S(q)= \frac{v_0^2}{2\Omega^2} \frac{q^2}{1+ b^2q^4/\Omega^2} = \frac{1}{2}(Rq)^2 + O(q^6). \label{30}
\end{equation}
This is nothing but hyperuniform scaling with exponent $\alpha =2$, which was reported in Ref.~\cite{Li2019}.
Notice that the low wavenumber behavior of $S(q)$ does not depend on the parameter $b$ and is characterized solely by the orbital radius $R=v_0/\Omega$.
Also, at length scale {s} smaller than $\sqrt{b/\Omega}$, \eq{30} behaves as $S(q) \sim q^{-2}$.
This means that the density fluctuations increase at the intermediate length scale.
That is to say, cABP in $D\rightarrow 0$ exhibits both increases and decreases in density fluctuations depending on the length scale: for $q > \sqrt{\Omega/b}$, the density fluctuations increase whereas for $q < \sqrt{\Omega/b}$, the system is hyperuniform. 
This crossover behavior from the large density fluctuations to hyperuniformity is consistent with what has been observed numerically \cite{Li2019} and experimentally~\cite{Zhang2022}.

Hyperuniformity in the limit $D\rightarrow 0$ can be understood as a result of the reduction of the number of excited modes by the chirality~\cite{ikeda2023}. 
To see this, we rewrite \eq{28} as
\begin{equation}
 S(\bm q,\omega) = \frac{D}{2}\qty[
 \frac{1}{D^2+(\omega-\Omega)^2} + \frac{1}{D^2+(\omega+\Omega)^2} \label{sqomega}
 ]S_\RM{eq}(\bm q,\omega),
\end{equation}
where 
\begin{equation}
S_\RM{eq}(\bm q,\omega) := \frac{v_0^2 q^2}{\omega^2 + b^2 q^4} \label{sqomega_eq}
\end{equation}
is the dynamical structure factor in the overdamped equilibrium system at the temperature $T=v_0^2/(2\mu)$. 
The spectrum in the equilibrium system, \eq{sqomega_eq}, has a peak at $\omega=0$ called the Rayleigh line~\cite{hansen}, which has a width of $bq^2$. 
This corresponds to the thermally excited modes. 
In the limit $D\rightarrow 0$, \eq{sqomega} is reduced to
\begin{equation}
 S(\bm q,\omega) =\frac{\pi}{2} S_\RM{eq}(\bm q,\omega) \qty[\delta(\omega-\Omega) + \delta(\omega+\Omega)]. \label{sqomegaHU}
\end{equation}
This means that in the limit $D\rightarrow 0$, the excited modes are only $\omega = \pm \Omega$. 
Namely, the chirality reduces the finite width spectrum (Rayleigh line) to only two line spectra {of} infinitesimal width. 
The static structure factor is obtained by integrating \eq{sqomegaHU} over $\omega$: $S(q) = S_\RM{eq}(\bm q,\omega = \Omega)/2 \sim q^2$. 
This explains why the density fluctuations are suppressed compared with the thermal equilibrium fluctuations. 

In the another extreme case, $\Omega\rightarrow 0$ with finite $D$, in which Eqs.~(\ref{1}) and (\ref{2}) are reduced to the standard ABP, and \eq{29} takes the Ornstein-Zernike form:
\begin{equation}
S(q) = \frac{S_0}{1+(\xi q)^2}. \label{31}
\end{equation}
Here we defined the correlation length $\xi:= \sqrt{b/D}$ and $S_0:= {v_0^2}/({2b D})=\rho T_\RM{act}\chi$.
\eq{31} means that in the homogeneous fluid state of ABP, the density fluctuations have the spatial correlation with the correlation length $\xi$ \cite{kuroda2023}.  
Also, it is known that the spatial velocity correlation in the fluid state of ABP has the Ornstein-Zernike form~\cite{kuroda2023, Szamel2021EPL, marconi2021}. 
This large density correlation is a consequence of the presence of excess excited modes by the activity. 
In fact, \eq{sqomega} in $\Omega\rightarrow 0$ leads to $S(\bm q,\omega) = DS_\RM{eq}(\bm q,\omega)/(\omega ^2 + D^2)$, meaning that there are additional excited modes to the thermal modes.
Note that in the equilibrium limit $D\rightarrow \infty$, \eq{29} becomes $S(q) = S_0 = \rho T_\RM{act}\chi$, which is the well-known expression {for} the low wavenumber limit of $S(q)$ in the equilibrium statistical mechanics \cite{hansen}. 

If both $\Omega$ and $D$ are finite, $S(q)$ given by \eq{29} at the low wavenumber behaves as
\begin{equation}
S(q)=\mathcal S_0 + \mathcal S_2 q^2 +O(q^4), \label{32}
\end{equation}
where 
\begin{align}
&\mathcal S_0 := \frac{v_0^2 D}{2b(\Omega^2 +D^2)},\ \ \ \ \ 
\mathcal S_2 := \frac{v_0^2 (\Omega^2 - D^2)}{2(\Omega^2 +D^2)^2}.
\label{33}
\end{align}
The behavior of $S(q)$ changes depending on the sign of $\mathcal S_2$. 
When $\Omega < D$, $\mathcal S_2$ is negative, and $S(q)$ increases toward constant value $\mathcal S_0$ in the limit $q\rightarrow 0$.
On the other hand, when $\Omega > D$, $\mathcal S_2$ is positive, and $S(q) \sim q^2$ if $\mathcal S_0$ is small or $q$ is large enough. 
Although when $D$ is finite, $S(q)$ remains finite in the low wavenumber limit, the density fluctuations are suppressed for a wide range of length scales.
This is also consistent with the numerical study in Ref.~\cite{Li2019}.
Also, one can predict how $\mathcal S_0$ vanishes in the limit $D\rightarrow 0$. 
From the first equation of \eq{33}, we have $\mathcal S_0\sim D$ for small $D$.
This scaling relation is consistent with the numerical data for small $D$ (see \ref{apeC}).

\begin{figure*}[t]
\centering
  \includegraphics[width=16cm]{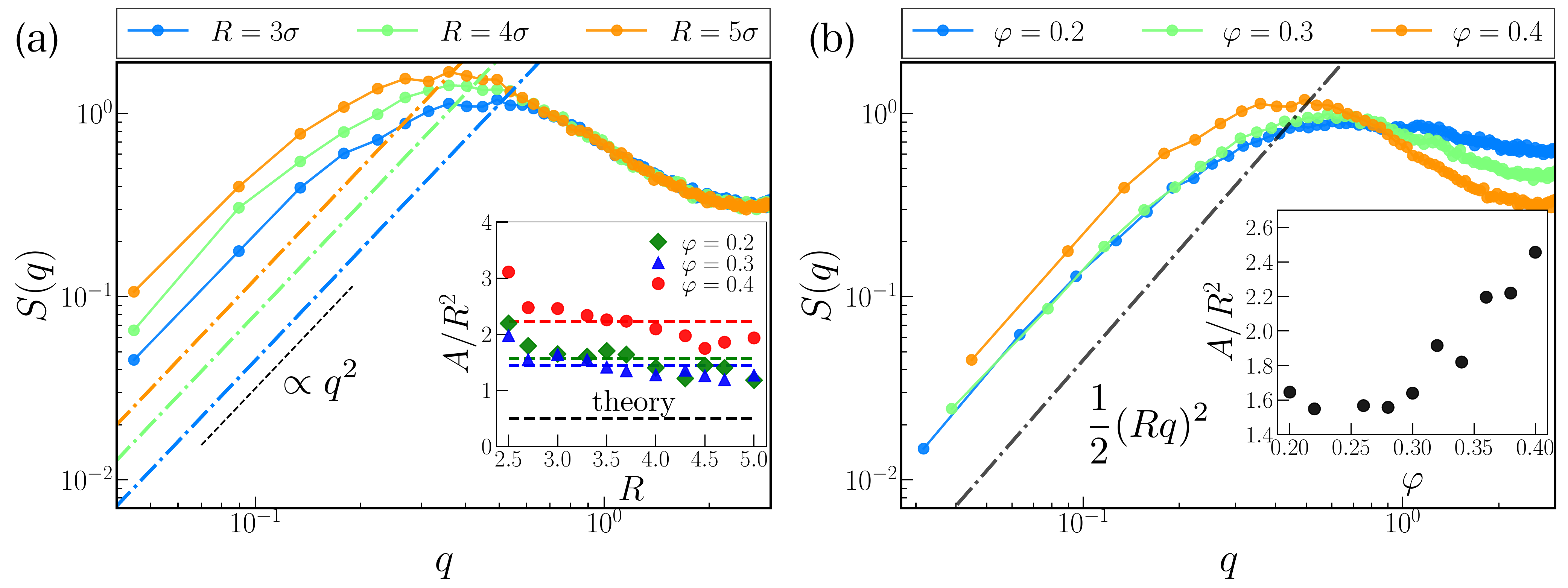}
 \caption{\label{fig1} Comparison of numerical data and theoretical results of the static structure factor $S(q)$ for various radius $R$ and packing fraction $\varphi$, at $D=0$. 
 Numerical data is shown by solid symbols. 
 (a) $S(q)$ for various $R$ at $\varphi =0.4$. The dot-dashed lines depict $S(q) = (Rq)^2/2$ predicted by the linearized theory, and the black dash line represents $q^2$ as a guide to the eye.
 $R$ dependence of the coefficient $A$ of $S(q) = Aq^2$ from numerical data are shown the inset. 
 The vertical axis is rescaled by $R^2$.
 The dotted line represents the average values. 
 (b) $S(q)$ for various $\varphi$ at $R=3\sigma$. The black dot-dashed line represents $S(q) = (Rq)^2/2$. 
 The coefficient $A$, obtained from the numerical data, as a function of $\varphi$ is shown in the inset.
 The vertical axis is rescaled by $R^2$.
}
\end{figure*}
\section{Comparison with numerical simulation}
\label{sim}
To assess the validity of our theory, we perform a numerical simulation.
The simulation setting is as follows.
We simulate Eqs.~(\ref{1}) and (\ref{2}) in the box $[0,L]^2$ and impose the periodic boundary condition.
We choose the Weeks--Chandler--Anderson (WCA) potential~\cite{WCA} as the pairwise potential:
\begin{equation}
U(r_{jk}) = 4\epsilon\left\{ \left( \frac{\sigma}{r_{jk} }\right)^{12} - \left( \frac{\sigma}{r_{jk} }\right)^{6} + \frac{1}{4} \right\} \theta( 2^{1/6}\sigma -r_{jk} ), \label{wca}
\end{equation}
where $r_{jk}=|\bm r_j-\bm r_k| $ is the distance between the particle $j$ and $k$, and $\sigma$ is the diameter of a particle. 
$\theta(\cdot)$ is the Heaviside step function, and $2^{1/6}\sigma$ is the cut-off length.
The unit of length scale, time scale, and energy are chosen $\sigma$, $\tau= \sigma/v_0$, and $\epsilon$, respectively. 
The number of particles is $N=1\times 10^4$. 
The controllable parameters are the packing fraction $\varphi=N\pi\sigma^2/(4L^2)$, {dimensionless} rotational diffusion constant $D\tau$, orbital radius $R/\sigma$, and energy ratio $\mu\epsilon/(\sigma v_0)$. Here we set $\mu\epsilon/(\sigma v_0)=1/24$ so that the repulsive force  and active force $v_0\bm e(\phi_j)$ are balanced at $r=\sigma$.
To numerically integrate the equation of motion, we use the Euler method with a time step $\Delta t = 10^{-3}\tau$. 
The above simulation setting is the same as  {in} Ref.~\cite{Li2019}. 
We compare our theory to the numerical simulation for several parameters {where} hyperuniformity was observed in Ref.~\cite{Li2019}, {\it i.e.}, we set $D\tau=0$ (see \ref{apeC} for the results at finite $D$). 
Hence the parameters we control in this simulation are $\varphi$ and $R$.
We compute the static structure factor $S(q)$ by taking the time average after confirming that the system {is} sufficiently relaxed to the stationary state by monitoring the potential energy.

We show the numerical data of the static structure factor $S(q)$ for $R = 3\sigma,\ 5\sigma$ and $10\sigma$, at fixed packing fraction $\varphi=0.4$, in Fig.~\ref{fig1}(a).
The filled circles represent the numerical data, and the dot-dashed lines depict $S(q) = (Rq)^2/2$ which is the theoretical prediction, \eq{30}.
Although hyperuniform exponent ($\alpha=2$) agrees with the numerical data, 
the prefactor $A:=S(q)/q^2$ substantially deviates from the theoretical prediction $A=R^2/2$.
In the inset of Fig.~\ref{fig1}(a), we show the coefficient $A$ divided by $R^2$, as a function of $R$, for several $\varphi$. 
$A$ was obtained by the fit of $S(q) = Aq^2$.
The fitting range was chosen as $q<0.1$. 
We measured $A$ only for $2.5 \sigma \leq R \leq 5\sigma$ because we cannot see the $q^2$ behavior for large $R>5\sigma$ due to the limitation of the system size. 
Also, at very large $R$, the system undergoes the clustering and is no longer homogeneous~\cite{Li2019}. 
The black dashed line stands for the theoretical prediction $A=R^2/2$. The numerical data are much larger than the theoretical prediction.
The colored dashed lines represent the average values. 
As $A/R^2$ is almost constant except at the smallest $R$, we conclude $R$ dependence of $A$ is $A\propto R^2$.
However, the coefficient $A$ depends on the number density or packing fraction.
In Fig.~\ref{fig1}(b), we depict $S(q)$ for $\varphi = 0.2,\ 0.3,$ and $0.4$ at $R=3\sigma$. 
The black dot-dashed line indicates the theoretical prediction $S(q) = (Rq)^2/2$. 
Obviously, the coefficient $A$ depends on $\varphi$, contrary to our theoretical prediction.
The inset of Fig.~\ref{fig1}(b) shows the coefficient $A/R^2$ as a function of $\varphi$. 
$A$ is almost constant at low densities, $\varphi \lesssim 0.3$ and increases for $\varphi\gtrsim 0.3$.
Since our theory neglects the non-linear terms which would be important at higher densities, it is naturally expected that the agreement with  {the} numerical simulation would improve at lower densities. 
However, $A/R^2 \sim 1.5$ for $\varphi \lesssim 0.3$ is approximately three times larger than the theoretical prediction $A/R^2 = 0.5$.
This implies that the non-linear effects are important even for $\varphi \lesssim 0.3$.
Note that lower densities are inaccessible in numerical simulation to investigate the chirality-induced hyperuniformity because the system falls into the absorbing state at low densities, and another type of hyperuniformity caused by the criticality would contribute~\cite{Li2019}.  
To explain the density dependence of the coefficient, it would be required to take the non-linear effects and higher order terms in the equation of state \eq{9} into account. 
These tasks are left for future work.

\section{Effective particle representation}
\label{eff}

\begin{figure*}[t]
\centering
  \includegraphics[width=8cm]{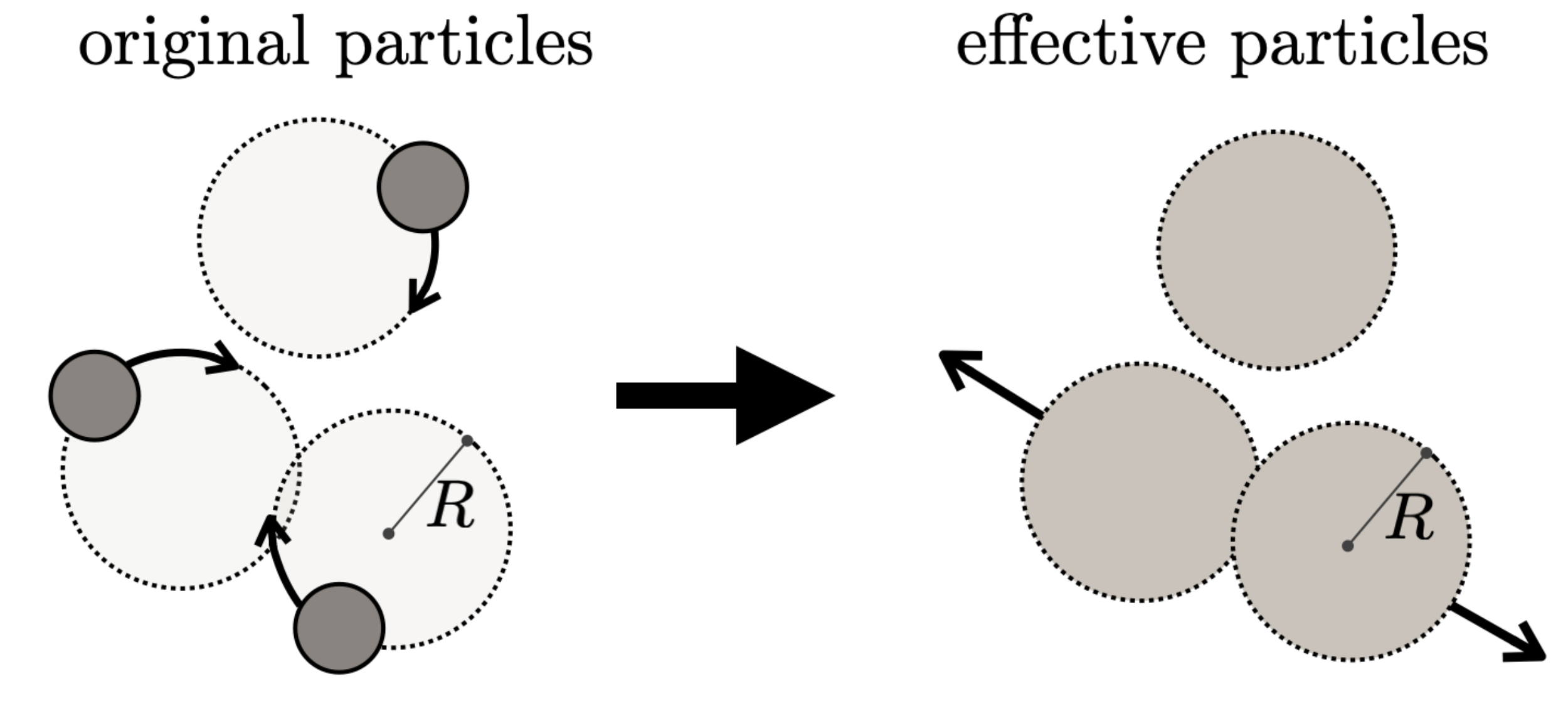}
 \caption{\label{fig2} Schematic of the effective particle representation. This picture corresponds to the case where the rotational diffusion constant is zero, at which each particle performs exactly circular motion without any noise.
 Each rotating trajectory can be regarded as an effective particle of the radius $R$.
}
\end{figure*}

In this section, we consider the effective particle representation {where} each rotating trajectory is regarded as a ``particle", as illustrated in Fig.~\ref{fig2}.
In Ref.~\cite{Li2019}, it is numerically shown that the center of mass of the effective particles also exhibits hyperuniformity characterized by $\alpha=2$, for the zero rotational diffusion constant ($D=0$). 
This hyperuniform exponent is the same as the one in the organization model with COMC~\cite{Hexner2017}. 
The dynamics of the effective particles resemble the random organization model: if the two effective particles overlap, then the two particles will undergo the ``kick" while conserving the center of mass of the particles. 
The isolated effective particles will stay in the same position~\cite{Hexner2017}.
Hexner {\it et al.}~\cite{Hexner2017} proposed that the density field in the random organization model, $\rho^\RM{o}(\bm r,t )$, is described by the following equation at large scales:
\begin{equation}
\p_t \rho^\RM{o}(\bm r,t ) = D_\RM{eff} \nabla^2\rho^\RM{o}(\bm r,t ) + \mathcal{B}\sqrt{\rho^\RM{o}}\nabla\cdot \bm \Upsilon(\bm r,t)
+ \mathcal{A}\sqrt{\rho^\RM{o}}\nabla^2 \eta(\bm r,t), \label{35}
\end{equation}
where $\bm \Upsilon(\bm r,t)$ and $ \eta(\bm r,t)$ are the Gaussian white noise.
$D_\RM{eff}$ is a diffusion coefficient, and $\mathcal A$ and $\mathcal B$ are the constants denoting the strength of the noises. 
For the random organization model, this equation with $\mathcal B=0$ has been derived from microscopic stochastic rule~\cite{Hexner2017}. 
The case of $\mathcal B=0$ corresponds to the athermal system.
The difference {between} \eq{35} {and} the standard diffusion equation is that
it contains a noise term with the square gradient $\nabla^2 \eta(\bm r,t)$ in addition to the standard conserved noise $\nabla\cdot \bm \Upsilon(\bm r,t)$. 
This term leads to the suppression of the density fluctuations~\cite{Li2019, Hexner2017}.
In fact, the static structure factor calculated from \eq{35} is given by
\begin{equation}
S^\RM{o}(q) := \frac{1}{N}\expval{\tilde{\rho}^\RM{o}(\bm q,0)\qty(\tilde{\rho}^\RM{o}(\bm q,0))^*} = \frac{\mathcal B^2}{2D_\RM{eff}} + \frac{\mathcal A^2}{2D_\RM{eff}}q^2, \label{So}
\end{equation}
which means hyperuniformity if $\mathcal B=0$. 
From the similarities between the chiral active fluid and the random organization model, the authors of Ref.~\cite{Li2019} suggest that the cABP in the effective particle representation should be governed by the same equation as \eq{35}.
In cABP, if the rotational diffusion constant $D$ is zero, the coefficient $\mathcal{B}$ should be zero because the noise term caused when $D$ is finite is given by the standard conserved noise. 
However, the microscopic derivation of \eq{35} for the chiral active particles is still absent.  
Here we show one of the routes to derive \eq{35} from the microscopic dynamics.
{In particular}, we show that the square gradient noise $\nabla^2 \eta(\bm r,t)$ is obtained from the coupling between the nematic tensor and the density field.

First, the center of mass of the effective particle $j$ is written as
\begin{equation}
\bm r_j^\RM{o}(t) = \bm r_j(t) + \frac{v_0}{\Omega^2}\bm \Omega \times \bm e(\phi_j(t)).
\label{36}
\end{equation}
Taking the time derivative of \eq{36} and using Eqs.~(\ref{1}) and (\ref{2}), we obtain 
\begin{equation}
\dv{\bm r_j^\RM{o}(t)}{t} = \mu\bm F_j^\RM{o}(t) + \sqrt{\frac{2v_0^2D}{\Omega^2}}\bm \xi^\RM{o}_j(t),
\end{equation}
where we defined
\begin{align}
&\bm F_j^\RM{o}(t):= \bm F_j(t) -\frac{v_0D}{\mu\Omega^2}\bm \Omega\times\bm e(\phi_j(t)), \\
&\bm \xi ^\RM{o} _j(t):= - \bm e(\phi_j(t))\bullet \eta_j(t).
\end{align}
Following the same procedure as  {for} the original particle representation shown in Sec.~\ref{FH}, we obtain the equation for the density field of the effective particles:
\begin{align}
\p_t\rho^\RM{o}(\bm r,t)=  D_\RM{eff}\nabla^2\rho^\RM{o}(\bm r,t) 
+ \sqrt{\frac{v_0^2D\rho^\RM{o}}{\Omega^2}}\nabla\cdot \bm \Upsilon(\bm r,t) 
+ \frac{v_0^2 D}{\Omega^2}\nabla\nabla :  \mathsf Q^\RM{o}(\bm r,t).
\label{40}
\end{align}
Here, we defined the diffusion coefficient for the effective particles as
\begin{equation}
D_\RM{eff} := \frac{\mu}{\rho^\RM{o}\chi^\RM{o}}  + \frac{v_0^2 D}{2\Omega^2},
\end{equation} 
where $\chi^\RM{o}$ is a constant that comes from the assumption for the pressure gradient (see \eq{9}).
\eq{40} contains the nematic tensor defined by
\begin{equation}
\mathsf Q^\RM{o}(\bm r,t):= \sum_{j=1}^{N}\qty(\bm e(\phi_j(t))\bm e(\phi_j(t))  - \frac{1}{2}\mathbbm{1})\delta(\bm r- \bm r_j^\RM{o}(t)), \label{42}
\end{equation}
where $\mathbbm 1$ is the identity matrix. The equation for the nematic tensor is described by 
\begin{align}
\p_t{\mathsf Q^\RM{o}(\bm r,t)}= -\nabla\cdot \qty(\frac{\bm J^\RM{o}(\bm r,t)\mathsf Q^\RM{o}(\bm r,t)}{\rho^\RM{o}(\bm r,t)})
 - 4D\mathsf Q^\RM{o}(\bm r,t) 
 + 2 \mathsf{\Omega}\mathsf Q^\RM{o}(\bm r,t) +  \sqrt{D\rho^\RM{o}(\bm r,t)}\mathsf \Theta(\bm r,t) .
 \label{43}
\end{align}
Here we used the assumption \eq{13} to derive the first term on the right-hand side.
The second-rank tensor $\mathsf \Theta(\bm r,t)$ is a Gaussian white noise which has zero mean and the correlation 
\begin{equation}
\expval{\Theta_{\alpha,\beta}(\bm r,t)\Theta_{\mu,\nu}(\bm r' ,t')} = \Delta_{\alpha\beta\mu\nu}\delta(\bm r-\bm r')\delta(t-t'), 
\end{equation}
where
\begin{equation}
\Delta_{\alpha\beta\mu\nu}:= -\delta_{\alpha,\beta}\delta_{\mu,\nu}
+\delta_{\alpha,\mu}\delta_{\beta,\nu}
+\delta_{\alpha,\nu}\delta_{\beta,\mu}. \label{delta}
\end{equation}
This noise term is obtained in a similar way  {as} described in \ref{apeB}.
To close the equations only for the density field, we exploit the adiabatic approximation~\cite{Cates_2013}. 
We assume that $\mathsf Q^\RM{o}(\bm r,t)$ is a fast variable since it has a relaxation time of $1/(4D)$.
Putting $\p_t\mathsf Q^\RM{o}(\bm r,t)\simeq \mathsf O$ and linearizing \eq{43}, we obtain 
\begin{align}
\p_t\delta\rho^\RM{o}(\bm r,t)=  D_\RM{eff}\nabla^2\delta\rho^\RM{o}(\bm r,t) 
+ \sqrt{\frac{v_0^2D\rho^\RM{o}}{\Omega^2}}\nabla\cdot \bm \Upsilon(\bm r,t) 
+ \sqrt{\rho^\RM{o}} \nabla\nabla : \mathsf A \mathsf \Theta(\bm r,t),
\label{46}
\end{align}
with 
\begin{equation}
\mathsf A :=\frac{v_0^2D^{3/2}}{2\Omega^2(4D^2+\Omega^2)}
\mqty(2D & -\Omega \\ \Omega & 2D).
\end{equation}
Using \eq{delta}, \eq{46} can be rewritten as 
\begin{equation}
\p_t\delta\rho^\RM{o}(\bm r,t)=  D_\RM{eff}\nabla^2\delta\rho^\RM{o}(\bm r,t) 
+\mathcal B \sqrt{\rho^\RM{o}}\nabla\cdot \bm \Upsilon(\bm r,t) 
+ \mathcal A\sqrt{\rho^\RM{o}} \nabla^2 \eta(\bm r,t),
\end{equation}
where 
\begin{equation}
\mathcal A:= \sqrt{\det[\mathsf A]} = \frac{v_0^2 D}{2\Omega^2}\sqrt{\frac{D}{4D^2+\Omega^2}}, 
\ \ \ \ \ \mathcal B := \sqrt{\frac{v_0^2 D}{\Omega^2}}, \label{AB}
\end{equation}
and $\eta(\bm r,t)$ is a Gaussian white noise.
This is the same form as \eq{35}. 
Thus {,} the static structure factor $S^\RM{o}(q)$ is given by \eq{So}.
As we mentioned above, the suppression term $q^2$ comes from the additional noise term $\nabla^2 \eta(\bm r,t)$.
In our derivation, this noise term stems from the coupling between the density field and the nematic tensor (the last term of \eq{40}).
Note that this coupling is forbidden in equilibrium systems because it cannot be expressed by the functional derivative with respect to the density field.
We remark that our theory cannot describe the exact hyperuniformity in the effective particle representation since  
the limit $D\rightarrow 0$ cannot be taken unlike $S(q)$ in the original particle representations, \eq{29}. 
Namely, $\mathcal B$ cannot be zero, so that $S^\RM{o}(q)$ remains the constant value in the limit $q\rightarrow 0$. 
This is because we have employed the adiabatic approximation which is valid only for finite $D$.
Derivation of a theory that is valid for the limit $D\rightarrow 0$ is left for future work.

\begin{figure*}[t]
\centering
  \includegraphics[width=10cm]{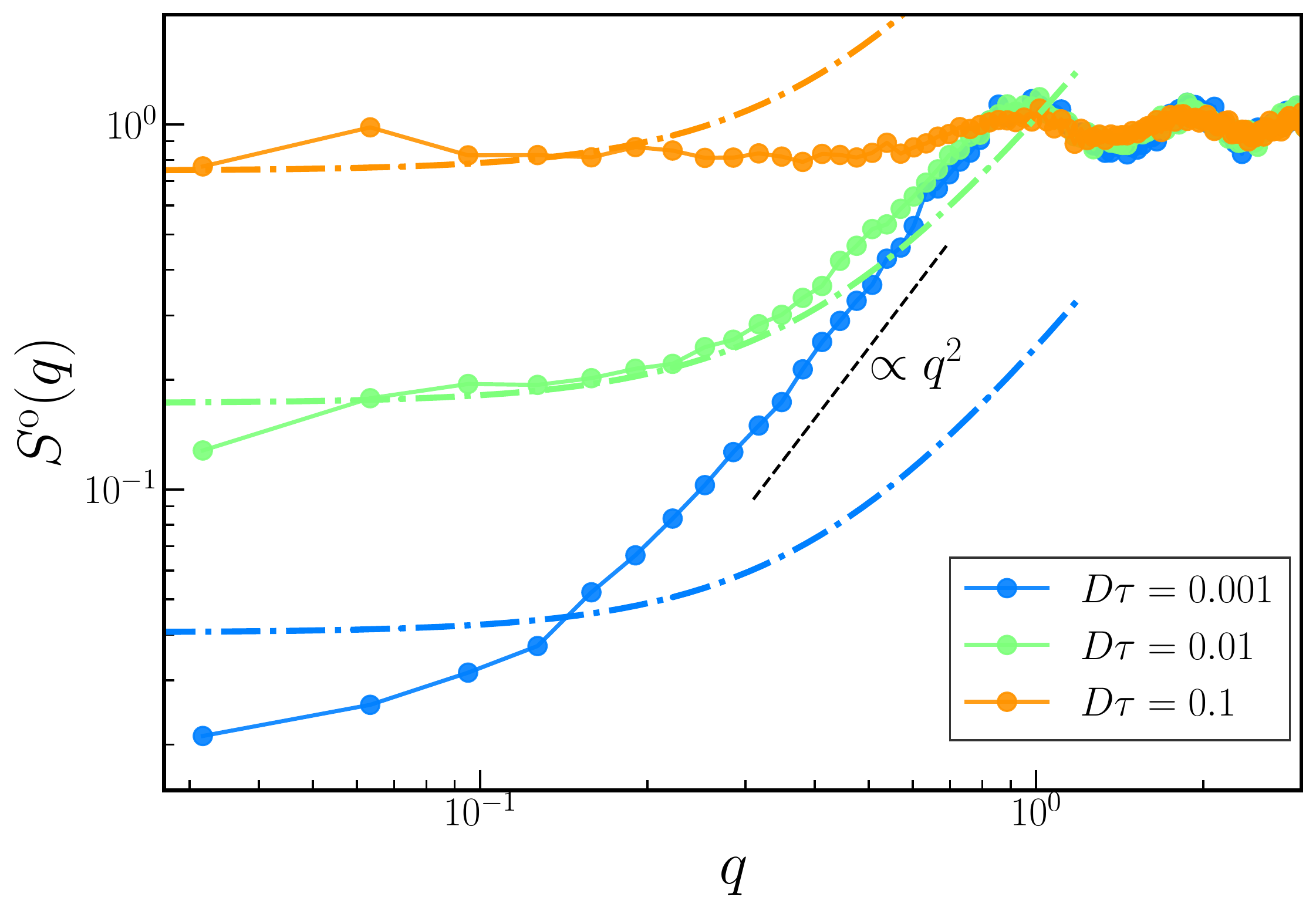}
 \caption{\label{fig3}  
 Comparison of theoretical predictions with numerical simulation in the effective particle representation. 
 The orbital radius and packing fraction are fixed as $R=3\sigma$ and $\varphi=0.2$, respectively. 
 The filled circle depict the numerically obtained $S^\RM{o}(q)$, and the dot-dashed lines are  {the} theoretical prediction by \eq{So}. $D_\RM{eff}$ is sole fitting parameter, and $\mathcal A$ and $\mathcal B$ is given by \eq{AB}.
 The back dashed line is $\propto q^2$ as a guide for the eyes.
}
\end{figure*}

Finally, we show the comparison of our theoretical prediction Eqs.~(\ref{So}) and (\ref{AB}) and numerical simulation at small but finite $D$, in Fig.~\ref{fig3}.
The filled circle in Fig.~\ref{fig3} is the numerical data of $S^\RM{o}(q)$. 
As reported in Ref.~\cite{Li2019}, the density fluctuations of the effective particles are suppressed, but $S^\RM{o}(q\rightarrow 0)$ does not go to zero and remains constant value because $D$ is finite.
We show the theoretical prediction, Eqs.~(\ref{So}) and (\ref{AB}), by the dot-dashed lines. 
$D_\RM{eff}$ in \eq{So} is sole fitting parameter.
The fitting range is $q<0.26$.
For the smallest rotational diffusion constant, $D\tau=0.001$, our theory fails the quantitative description, but for $D\tau=0.01$ and $0.1$, the theoretical lines agree with the numerical data. 
The reason why our theory fails for very small $D$ is supposed to be that we employed the adiabatic approximation to obtain the square gradient noise term.

\section{Conclusion and remarks}
\label{conc}
In this paper, we theoretically studied the density fluctuations in the two-dimensional chiral active fluid. 
We provided an effective fluctuating hydrodynamic description for the homogeneous fluid state of cABP where each particle has constant torque $\Omega$ and undergoes rotational diffusion.
The obtained hydrodynamic equations yield hyperuniformity $S(q)\propto q^2$ in the limit of the infinitesimal rotational diffusion constant $(D\rightarrow0)$.
This is consistent with the numerical observation in Ref.~\cite{Li2019}.
When the rotational diffusion constant is finite ($D\neq 0$), $S(q)$ remains constant in the low wavenumber limit.
However, in  {the} case $\Omega \gg D$, $S(q)$ is suppressed for a wide range of wavenumber. 
On the contrary, in case $\Omega <D$, $S(q)$ increases as $q$ {decreases}.
Also, our theoretical expression of $S(q)$ explains a large density fluctuation at intermediate length scales and its crossover to hyperuniformity, which had been observed in numerical \cite{Li2019} and experimental studies \cite{Zhang2022}. 
We also performed the numerical simulation for cABP at $D=0$ and compared the numerical data with theoretical results. 
Our theory can yield hyperuniformity characterized by the same exponent as the numerical data
but fails to derive the coefficient quantitatively.
In our theory, the coefficient is {determined} only  {by} the orbital radius of each particle, $R=\Omega/v_0$.
The $R$ dependence of the coefficient almost agrees with the numerical data.
However, numerical results showed that the coefficient depends on also the number density. 
A complete quantitative description of hyperuniformity is an important future task.
We also theoretically considered the effective particle representation, in which each rotating trajectory of a particle is regarded as an effective particle, as suggested in Ref.~\cite{Li2019}.
In this representation, there is a conjecture that the dynamics of the density field {is} driven by the square gradient noise in addition to the standard conserved noise~\cite{Li2019}.
Here we justified this conjecture from the microscopic dynamics, at least in the case of finite $D$. 
That is, the square gradient noise term was derived for cABP.
In cABP, this noise term stems from a nonequilibrium coupling between the density field and the nematic tensor. 
However, our hydrodynamic description of the effective particle representation is not complete: the exact hyperuniformity cannot be derived, and for the small rotational diffusion constant, the derived $S(q)$ agrees with the numerical result only qualitatively, since our theory is valid only for finite $D$.
{Further studies} would be required to derive the exact coarse-grained description for the effective particle representation in future work. 

Before concluding this paper, we make several remarks.
First, we address that the hypruniformity $S(q) \propto q^2$ {was} obtained by starting from the fluctuating hydrodynamic equations with the finite $D$ and taking $D\rightarrow 0$ after {calculating} the correlation function. 
Deriving the same result from the equation with $D=0$ from the outset is not easy because it requires deriving the fluctuating macroscopic description from a deterministic process.
That is a hard task even in passive systems.
Second, our theory is valid only for the homogeneous phase and cannot explain the MIPS or dynamical clustering in cABP that is reported in several studies \cite{Li2019, Bickmann2022, Ma2022}.
{In particular}, Z. Ma and R. Ni \cite{Ma2022} reported that there are two unstable modes, one{corresponding} to the occurrence of conventional MIPS and the other to the disintegration of MIPS.
These instabilities are the result of a negative diffusion constant caused by the decreases in the self-propelling as density increases. 
Our theory cannot capture these instabilities because several terms are missing in Eqs.~(\ref{20}) and (\ref{21}), including the density dependence of the self-propelling speed, which exists in the mean-field description for such clustering~\cite{Li2019, Bickmann2022, Ma2022}. 
Constructing a unified theory including these instabilities is important for future work.
Third, we completely neglected the hydrodynamic interaction, while several studies suggested hyperuniformity induced by the hydrodynamic interaction~\cite{Huang2021, Oppenheimer:2022aa}.
Considering the effect of the hydrodynamic interaction in chiral active fluids would be an important direction for future work.
Although {the above issues remain to be fully understood}, {our simple theory will serve as} the first step toward understanding the dynamical hyperuniformity in chiral active fluids.
\ack
We thank Harukuni Ikeda, Daiki Nishiguchi, Kazumasa A. Takeuchi, and Kyosuke Adachi for fruitful discussions. 
This work was supported by KAKENHI (Grant Number JP20H00128, JP23KJ1068), and JST SPRING (Grant Number JPMJSP2125).

\appendix
\section{Free motion of a chiral active Brownian particle}
\label{apeA}
We consider Eqs.(\ref{1}) and (\ref{2}) with $\bm F_j = \bm 0$.
In this appendix, we review the free motion of the single chiral active Brownian particle. 
In particular, orientational time correlations, mean displacement, and mean square displacement are calculated.
Note that, here, we omit the particle index $j$.

\subsection{Orientational time correlation}
First, we calculate the autocorrelation function of the orientational vector $\bm e(\phi(t))$.
To this aim, it is convenient to use the characteristic function for the Gaussian process. 
Let $\{X(t_1), X(t_2),..., X(t_n)\}$ be any sequence of points of the Gaussian process $X(t)$. The characteristic function is given by
\begin{align}
&\expval{\exp[i\sum_{a=1}^{n}\kappa_a X(t_a) ] } =\exp[ -\frac{1}{2}\sum_{a,b=1}^{n}\kappa_a \kappa_b C_\RM{X}(t_a,t_b) + i \sum_{a=1}^{n} \kappa_a\expval{X(t_a)}],
\label{a1}
\end{align}
where $C_\RM{X}(t,t')$ is the covariance between $X(t)$ and $X(t')$ defined by
\begin{equation}
C_\RM{X}(t,t') = \expval{X(t)X(t')}  - \expval{X(t)}\expval{X(t')}.
\end{equation}
From \eq{2}, the angle $\phi(t)$ is a Gaussian process characterized by the expectation value 
\begin{equation}
\expval{\phi(t)}= \phi(0) + \Omega t
\end{equation}
and the covariance between two times
\begin{equation}
 C_\phi(t,t') = 2Dt\wedge t'.
\end{equation}
Here, $t\wedge t'$ represents the smaller of $t$ and $t'$.

By expressing $\cos\phi(t)$ and $\sin\phi(t)$ in terms of $e^{\pm i\phi(t)}$ and using \eq{a1} at $n=1$,  we get 
\begin{align}
&\expval{\cos\phi(t)}= \cos(\phi(0) + \Omega t) e^{-Dt}, \label{a5} \\
&\expval{\sin\phi(t)}= \sin(\phi(0) + \Omega t) e^{-Dt}. \label{a6}
\end{align}
Likewise, \eq{a1} at $n=2$ leads to 
\begin{align}
\expval{e^{\pm i( \phi(t) + \phi(t'))}}
&= e^{-D(t + t' + 2 t\wedge t') \pm i\qty{2\phi(0) + \Omega(t+t')} }, \\
\expval{e^{\pm i( \phi(t) - \phi(t'))}}
&= e^{-D|t-t'| \pm i\Omega(t-t') }.
\end{align}
We thus obtain the orientational time correlations as
\begin{align}
&\expval{\cos\phi(t)\cos\phi(t')}  = \frac{1}{2}\Bigl\{
\cos[\Omega(t-t')]e^{-D|t-t'|} 
+ \cos[2\phi(0) + \Omega(t+t')]e^{-D(t+t'+2t\wedge t')}
\Bigr\}, \label{a9}\\
&\expval{\sin\phi(t)\sin\phi(t')}
=\frac{1}{2}\Bigl\{
\cos[\Omega(t-t')]e^{-D|t-t'|} 
- \cos[2\phi(0) + \Omega(t+t')]e^{-D(t+t'+2t\wedge t')}
\Bigr\}, \label{a10}\\
&\expval{\cos\phi(t)\sin\phi(t')}
= \frac{1}{2}\Bigl\{
-\sin[\Omega(t-t')] e^{-D|t-t'|}  
+ \sin[2\phi(0) + \Omega(t+t') ]e^{-D(t+t'+2t\wedge t')}
\Bigr\}.\label{a11}
\end{align}
Note that the term $\bm \xi_j(t):=v_0\bm e(\phi_j)$ in \eq{1} can be regarded as a colored noise which behaves as following for large $t$ and $t'$:
\begin{equation}
\expval{\xi^{(\alpha)}_j(t)\xi^{(\alpha)}_k(t')} \simeq \frac{v_0^2}{2} \cos[\Omega(t-t')]e^{-D\abs{t-t'}} \delta_{j,k}
\end{equation}
for $\alpha =x,y$ and 
\begin{equation}
\expval{\xi^{(\alpha)}_j(t)\xi^{(\beta)}_k(t')} \simeq \epsilon_{\alpha,\beta}\frac{v_0^2}{2} \sin[\Omega(t-t')]e^{-D\abs{t-t'}} \delta_{j,k}
\end{equation}
for $\alpha \neq \beta$.
Here $\epsilon_{x,y}=-1$ and $\epsilon_{y,x}=1$.
In the limit $D\rightarrow \infty$ while keeping $T_\RM{act}=v_0^2/(2 D \mu)$ constant, $\bm \xi_j(t)$ is reduced to a Gaussian white noise satisfying the fluctuation-dissipation relation 
\begin{equation}
\expval{\xi^{(\alpha)}_j(t)\xi^{(\beta)}_k(t')} =2\mu T_\RM{act}\delta_{\alpha,\beta}\delta_{j,k} \delta(t-t').
\end{equation}

\subsection{Mean displacement and mean square displacement}
Next, we calculate the mean displacement and mean square displacement. 
We write the displacement as $\Delta\bm r(t) = \bm r(t) - \bm r(0)$. 
The $x$ and $y$ components of $\Delta\bm r(t)$ are obtained from Eqs.~(\ref{a5}) and (\ref{a6}):
\begin{align}
&\expval{\Delta x(t)} = v_0\int_{0}^{t}\dd s\ \expval{\cos\phi(t)} \\
&= \frac{v_0}{D^2 + \Omega^2}\bigl[ \bigl\{ \Omega \sin(\phi(0) + \Omega t) \notag - D\cos(\phi(0) + \Omega t) \bigr\}e^{-Dt}  
- \Omega \sin\phi(0) + D\cos\phi(0) \bigr],  \\
&\expval{\Delta y(t)}= v_0\int_{0}^{t}\dd s\ \expval{\sin\phi(t)}  \\
&= \frac{v_0}{D^2 + \Omega^2}\bigl[  -\bigl\{ \Omega \cos(\phi(0) + \Omega t) + D\sin(\phi(0) + \Omega t) \bigr\} e^{-Dt}
+ \Omega \cos\phi(0) + D\sin\phi(0) \bigr]. \notag 
\end{align}
The mean squared displacement $\expval*{\abs{\Delta \bm r(t)}^2}$ is obtained from the orientational time correlation functions.
Using Eqs.~(\ref{a9}) and (\ref{a10}), we have
\begin{align}
&\expval{\abs{\Delta \bm r(t)}^2}  = v_0^2 \int_{0}^{t}\dd s\int_{0}^{t}\dd s'\expval{\cos\phi(s)\cos\phi(s')+\sin\phi(s)\sin\phi(s')} \notag\\
&=\frac{2v_0^2}{(D^2+\Omega^2)^2}\bigl[(\Omega^2-D^2)+D(D^2+\Omega^2)t 
+e^{-Dt} \bigl\{(D^2 - \Omega^2 )\cos(\Omega t) - 2D \Omega \sin(\Omega t) \bigr\} \bigr].
\end{align}
In the long time limit, the behavior of a particle is diffusive with the $\Omega$ dependent diffusion coefficient defined by
\begin{equation}
D_\RM{\Omega} := \lim_{t\rightarrow \infty}\frac{\expval*{\abs{\Delta \bm r(t)}^2}}{4t} = \frac{D_\RM{act}}{1+(\Omega/D)^2}.
\end{equation}
Here, $D_\RM{act}:= v_0^2/(2D)$ is the diffusion coefficient of the standard active Brownian particle~\cite{Bechinger2016RMP}.
\section{Derivation of \eq{18}}

\label{apeB}
In this appendix, we show that the noise term $\bm \Lambda(\bm r,t)$, \eq{17}, is statistically equivalent to the Gaussian white noise, \eq{18}, in the large $N$ limit.

First, since the definition of $\bm \Lambda(\bm r,t)$, \eq{17}, is expressed by the It\^o representation, 
obviously $\expval{\bm \Lambda(\bm r,t)}=\bm 0$.
Next, we evaluate the correlation function $\expval{\bm \Lambda(\bm r,t)\bm \Lambda(\bm r',t')}$.
The $(x,x)$-component of this correlation is given by
\begin{align}
&\expval{\Lambda_x(\bm r,t)\Lambda_x(\bm r',t')} 
=2D\sum_{j=1}^{N}\expval{\sin^2(\phi_j(t) )}\delta(\bm r-\bm r_j(t)) \delta(\bm r-\bm r')\delta(t-t'). 
\label{b1}
\end{align}
Here, we used $\expval{\eta_j(t)\eta_k(t')} = \delta_{j,k}\delta(t-t')$ and some properties of the delta function. 
From \eq{a10}, the expectation value of $\sin^2(\phi_j(t))$ is 
\begin{equation}
\expval{\sin^2(\phi_j(t))}
= \frac{1}{2} + \frac{1}{2}\cos(2(\phi _j(0) + \Omega t))e^{-4Dt}.
\end{equation}
Thus, \eq{b1} becomes 
\begin{align}
&\expval{\Lambda_x(\bm r,t)\Lambda_x(\bm r',t')} \notag\\
&=D\delta(\bm r-\bm r')\delta(t-t') \Bigl[ \rho(\bm r,t)  
+ \qty{ \mathcal C_N(\bm r,t) \cos(2\Omega t) - \mathcal S_N(\bm r,t)\sin(2\Omega t) }e^{-4Dt}  \Bigr], \label{b3}
\end{align}
where we defined 
\begin{align}
&\mathcal C_N(\bm r,t) := \sum_{j=1}^{N}\cos(2\phi_j(0))\delta(\bm r-\bm r_j(t)) \label{b4}, \\
&\mathcal S_N(\bm r,t) := \sum_{j=1}^{N}\sin(2\phi_j(0))\delta(\bm r-\bm r_j(t)). \label{b5}
\end{align}
Now we evaluate $\mathcal C_N(\bm r,t) $ and $\mathcal S_N(\bm r,t)$.
We assume that the initial value $\phi_j(0)$ is given completely random by the uniform distribution on the interval $[0,2\pi)$.
Then $\cos(2\phi_j(0))$ and $\sin(2\phi_j(0))$ take random values given by the uniform distribution on the interval $[-1,1]$.
From the low of large numbers, the following equations hold with probability 1:
\begin{align}
&\lim_{N\rightarrow \infty}\frac{1}{N}{\sum_{j=1}^N\cos(2\phi_j(0))}=0, \label{b6}\\
& \lim_{N\rightarrow \infty}\frac{1}{N} {\sum_{j=1}^N\sin(2\phi_j(0))}=0. \label{b7}
\end{align}
Here we assume the system has the volume $V= L^2$ with the periodic boundary condition. 
The Fourier series expansion of \eq{b4} yields
\begin{equation}
\mathcal C_N(\bm r,t) = \sum_{\bm n \in\mathbb{Z}^2}\tilde{ \mathcal C}_N(\bm n,t)e^{2\pi i\bm n\cdot \bm r/L},
\end{equation}
with 
\begin{align}
\tilde{ \mathcal C}_N(\bm n,t) &:= \frac{1}{V}\int_{V}\dd[2]\bm r\ \mathcal C_N(\bm r,t)e^{-2\pi i\bm n\cdot \bm r/L} = \frac{\rho}{N}\sum_{j=1}^{N}\cos(2\phi_j(0))e^{-2\pi i\bm n\cdot \bm r_j(t)/L}. \label{b9}
\end{align}
Since $\sum_{j=1}^{N}\cos(2\phi_j(0))=o(N)$ from \eq{b6}, we have the evaluation for the Fourier coefficient \eq{b9}:
\begin{equation}
\abs{\tilde{ \mathcal C}_N(\bm q,t)} \leq \frac{\rho}{N}\abs{\sum_{j=1}^{N}\cos(2\phi_j(0))} = o(1).
\end{equation}
Thus, we obtain
\begin{equation}
\lim_{N\rightarrow \infty}\mathcal C_N(\bm r,t) =0. \label{b11}
\end{equation}
In the same way, \eq{b5} turns out to be $0$ in the large $N$ limit:
\begin{equation}
\lim_{N\rightarrow \infty}\mathcal S_N(\bm r,t) =0. \label{b12}
\end{equation}
Combining Eqs. (\ref{b11}) and (\ref{b12}), in the limit of $N\rightarrow \infty$, \eq{b3} becomes
\begin{equation}
\expval{\Lambda_x(\bm r,t)\Lambda_x(\bm r',t')} =D\rho(\bm r,t)\delta(\bm r-\bm r')\delta(t-t') + o(1).
\label{b13}
\end{equation}
Similarly, for the $(y,y)$-component, we have 
\begin{equation}
\expval{\Lambda_y(\bm r,t)\Lambda_y(\bm r',t')} =D\rho(\bm r,t)\delta(\bm r-\bm r')\delta(t-t') + o(1).
\label{b14}
\end{equation}
The cross correlation between $(x,y)$-component becomes $o(1)$ because the equation
\begin{equation}
\expval{\cos\phi_j(t)\sin\phi_j(t)} = \frac{1}{2}\sin(2(\phi_j(0) +\Omega t))e^{-4Dt}
\end{equation}
 leads to
 \begin{align}
\expval{\Lambda_x(\bm r,t)\Lambda_y(\bm r',t')} &=D\delta(\bm r-\bm r')\delta(t-t')
 \bigl( \mathcal C_N(\bm r,t) \sin(2\Omega t) 
+ \mathcal S_N(\bm r,t)\cos(2\Omega t) \bigr )e^{-4Dt}   \notag \\
&=o(1),
\label{b16}
\end{align}
in the large $N$ limit.
Collecting Eqs.~(\ref{b13}), (\ref{b14}), and (\ref{b16}), we reach the desired equation.
\section{Equilibrium limit of the nonlinear equations}
\label{Eqlim}
We summarize the nonlinear equations obtained in Sec.~\ref{FH} for convenience:
\begin{align}
\p_t{\rho(\bm r,t)} &= -\nabla\cdot \bm J(\bm r,t), \label{Eq1} \\
\bm J(\bm r,t)  &= - \mu \rho(\bm r,t)\nabla\fdv{\mathcal F[\rho(\cdot,t)]}{\rho(\bm r,t)}+ v_0\bm p(\bm r,t),\label{Eq2}\\
\p_t\bm p(\bm r,t) &= -\nabla\cdot\qty(\frac{\bm J(\bm r,t)\bm p(\bm r,t)}{\rho(\bm r,t)})
- D\bm p(\bm r,t) 
+ \bm \Omega \times \bm p(\bm r,t) + \sqrt{D\rho(\bm r,t)}\bm \Upsilon(\bm r,t).\label{Eq3}
\end{align}
The functional $\mathcal F[\rho]$ is given by \eq{free_ene}.
In this appendix, we show that Eqs.~(\ref{Eq1})-(\ref{Eq3}) are reduced to the standard Dean--Kawasaki equation~\cite{Dean_1996, Kawasaki1994} in the limit $D\rightarrow \infty$~\cite{Nakamura_2009}.

\eq{Eq1} can be rewritten as 
\begin{align}
\rho(\bm r,t +\dd t) - \rho(\bm r,t )  &= - \nabla  \cdot \int_{t}^{t+\dd t}\dd s\ \bm J(\bm r,s) \notag \\
&=  - \nabla  \cdot \int_{t}^{t+\dd t}\dd s\ \qty[ -\mu \rho(\bm r,s)\nabla\fdv{\mathcal F[\rho(\cdot,s)]}{\rho(\bm r,s)}+ v_0\bm p(\bm r,s)
].
\end{align}
In the following, we evaluate the integral $\int_t^{t+\dd t}\dd s\ \bm p(\bm r,s)$.
\eq{Eq3} can be formally solved as
\begin{equation}
\bm p(\bm r,t) =  - \nabla\cdot \mathsf C(\bm r,t) + \bm \chi(\bm r,t),
\end{equation}
where
\begin{equation}
\mathsf C(\bm r,t)  := \int_{-\infty}^{t} \dd s\ \mathsf R(\Omega (t-s))e^{- D (t-s)} \frac{\bm J(\bm r,s)\bm p(\bm r,s)}{\rho(\bm r,s)}, \label{Eq6}
\end{equation}
\begin{equation}
 \bm \chi(\bm r,t) := \int_{-\infty}^{t} \sqrt{D\rho(\bm r,s)}\mathsf R(\Omega (t-s))e^{- D (t-s)} \dd \bm W(\bm r,s). \label{Eq7}
\end{equation}
$\bm W(\bm r,t)$ is the Wiener process, and the matrix $\mathsf R(x)$ is defined by
\begin{equation}
\mathsf R(x) := \mqty( \cos x & -\sin x \\ \sin x & \cos x).
\end{equation}
First, we show that \eq{Eq7} is reduced to a Gaussian white noise in the limit $D\rightarrow \infty$.
Using the relation~\cite{gardiner}
\begin{equation}
\int_{t_0}^{t} \dd W(s)F(s)\int_{t_0}^{t'} \dd W(s')G(s') = \int_{t_0}^{t\wedge t'} \dd s\ F(s)G(s), \label{Eq9}
\end{equation}
we have
\begin{align}
\expval{\chi_\alpha(\bm r,t) \chi_\beta(\bm r',t') }
&= DR_{\alpha,\beta}(\Omega(t-t'))\delta_{\alpha,\beta}\delta(\bm r-\bm r') \int^{t\wedge t'}_{-\infty}\dd s\ \rho(\bm r,s) e^{-D(t+t' + 2s)} \notag \\
&=DR_{\alpha,\beta}(\Omega(t-t')) e^{-D|t-t'|}\delta_{\alpha,\beta}\delta(\bm r-\bm r') \int^{t\wedge t'}_{-\infty}\dd s\ \rho(\bm r,s) e^{-2D(s-t\wedge t')} .
\label{Eq10}
\end{align}
At the last equality, we used $t+t' = |t-t'| + 2t\wedge t'$.
To consider the limit $D\rightarrow \infty$, we employ the formal expansion 
\begin{equation}
\frac{D}{2}e^{-D|t|} = \sum_{n=0}^{\infty} \frac{1}{D^{2n}}\dv[2n]{t}\delta(t). \label{Eq11}
\end{equation}
Substituting \eq{Eq11} into \eq{Eq10}, we get 
\begin{equation}
\expval{\chi_\alpha(\bm r,t) \chi_\beta(\bm r',t') } = \frac{\rho(\bm r,t)}{D}\delta_{\alpha,\beta}\delta(\bm r-\bm r')\delta(t-t') + O(D^{-3})\ \ \ \ (D\rightarrow \infty).
\end{equation}
Next, we investigate \eq{Eq6} in the limit $D\rightarrow \infty$. 
Noting that $\Upsilon_\alpha(\bm r,t) \sim O(\dd t^{-1/2})$, the integral of $\mathsf C(\bm r,t)$ in an  infinitesimal interval is evaluated as 
\begin{align}
&\int_{t}^{t+\dd t} \dd s\ C_{\alpha,\beta}(\bm r,s) 
= \int_{t}^{t+\dd t} \dd s\int_{-\infty}^{s}\dd s'\ R_{\alpha,\beta}(\Omega (s-s'))e^{-D(s-s')}\frac{J_\alpha(\bm r,s')p_\beta(\bm r,s')}{\rho(\bm r,s')} \notag \\
&= \int_{t}^{t+\dd t} \dd s\int_{-\infty}^{s}\dd s'\ v_0R_{\alpha,\beta}(\Omega (s-s'))e^{-D(s-s')}\frac{\chi_\alpha(\bm r,s')\chi_\beta(\bm r,s')}{\rho(\bm r,s')} + o(\dd t) \notag \\
&=\frac{v_0}{D} \int_{t}^{t+\dd t} \dd s\ \frac{\chi_\alpha(\bm r,s)\chi_\beta(\bm r,s)}{\rho(\bm r,s)} +O(D^{-3}) +  o(\dd t).
\end{align}
Recall that in the derivation of Eqs.~(\ref{Eq1})-(\ref{Eq3}), we assumed \eq{13}.
This leads to the equality 
$\rho(\bm r,t)\rho(\bm r,t) = \rho(\bm r,t)\delta(\bm r-\bm r)$.
Note that $\delta(\bm r-\bm r)$ should be defined by using a proper cutoff length scale.
Using this equality and \eq{Eq9}, we obtain the following relation:
\begin{align}
\chi_\alpha(\bm r, t)\chi_\beta(\bm r, t) &= D\delta_{\alpha,\beta}\int_{-\infty}^{t}\dd s\ \rho^2(\bm r,s)e^{-2D(t-s)} \notag \\
&= \frac{1}{2}\delta_{\alpha,\beta}\rho^2(\bm r,t) + O(D^{-2}) .
\end{align}
Thus, we have 
\begin{equation}
\int_{t}^{t+\dd t} \dd s\ C_{\alpha,\beta}(\bm r,s) =\frac{v_0}{2D}\delta_{\alpha,\beta} \rho(\bm r,t)\dd t+ O(D^{-3}) + o(\dd t).
\end{equation}
Collecting the above results, we get
\begin{equation}
\int_{t}^{t+\dd t}\dd s\ v_0\bm p(\bm r,t) = \frac{v_0^2}{2D} \nabla \rho(\bm r,t)\dd t + \sqrt{\frac{v_0^2\rho(\bm r,t)}{D}}\bm \xi(\bm r,t) \dd t + O(D^{-3}) + o(\dd t), \label{Eq16}
\end{equation}
where $\bm \xi(\bm r,t)$ is a Gaussian white noise satisfying $\expval{\bm \xi(\bm r,t)}=\bm 0$ and $\expval{\xi_\alpha(\bm r,t) \xi_\beta(\bm r,t)} = \delta_{\alpha,\beta}\delta(\bm r-\bm r')\delta(t-t') $.
In the limit $D\rightarrow \infty$ while keeping $ T_\RM{act} = v_0^2/(2D\mu)$ constant, \eq{16} becomes 
\begin{equation}
\int_{t}^{t+\dd t}\dd s\ v_0\bm p(\bm r,t) = \mu T_\RM{act} \nabla \rho(\bm r,t)\dd t + \sqrt{2\mu T_\RM{act} \rho(\bm r,t)}\bm \xi(\bm r,t) \dd t + o(\dd t). \label{Eq17}
\end{equation}
The diffusion term of \eq{Eq17} can be written as
\begin{equation}
T_\RM{act} \nabla \rho(\bm r,t) = \rho(\bm r,t) \nabla\fdv{\mathcal F^{\RM{id}}[\rho(\cdot,t)]}{\rho(\bm r,t)},
\end{equation}
where $\mathcal F^{\RM{id}}[\rho(\cdot,t)]$ is the ideal part of the ``free energy" functional:
\begin{equation}
\mathcal F^{\RM{id}}[\rho(\cdot,t)] := T_\RM{act}\int_{V}\dd[2]\bm r\ \rho(\bm r,t)(\log \rho(\bm r,t) -1).
\end{equation}
Hence, in the limit $D\rightarrow \infty$ with fixed  $T_\RM{act} = v_0^2/(2D\mu)$, Eqs.(\ref{Eq1})-(\ref{Eq3}) are reduced to the Dean--Kawasaki equation~\cite{Dean_1996, Kawasaki1994}:
\begin{align}
\p_t{\rho(\bm r,t)} &= -\nabla\cdot \bm J(\bm r,t), \\
\bm J(\bm r,t)  &= - \mu \rho(\bm r,t)\nabla\fdv{\mathfrak F[\rho(\cdot,t)]}{\rho(\bm r,t)}+\sqrt{2\mu T_\RM{act} \rho(\bm r,t)}\bm \xi(\bm r,t) ,\\
\mathfrak F[\rho(\cdot,t)] &:= \mathcal F^{\RM{id}}[\rho(\cdot,t)] + \mathcal F[\rho(\cdot,t)] .
\end{align}
\section{Numerical results at finite $D$}
\label{apeC}
\begin{figure*}[t]
\centering
  \includegraphics[width=16cm]{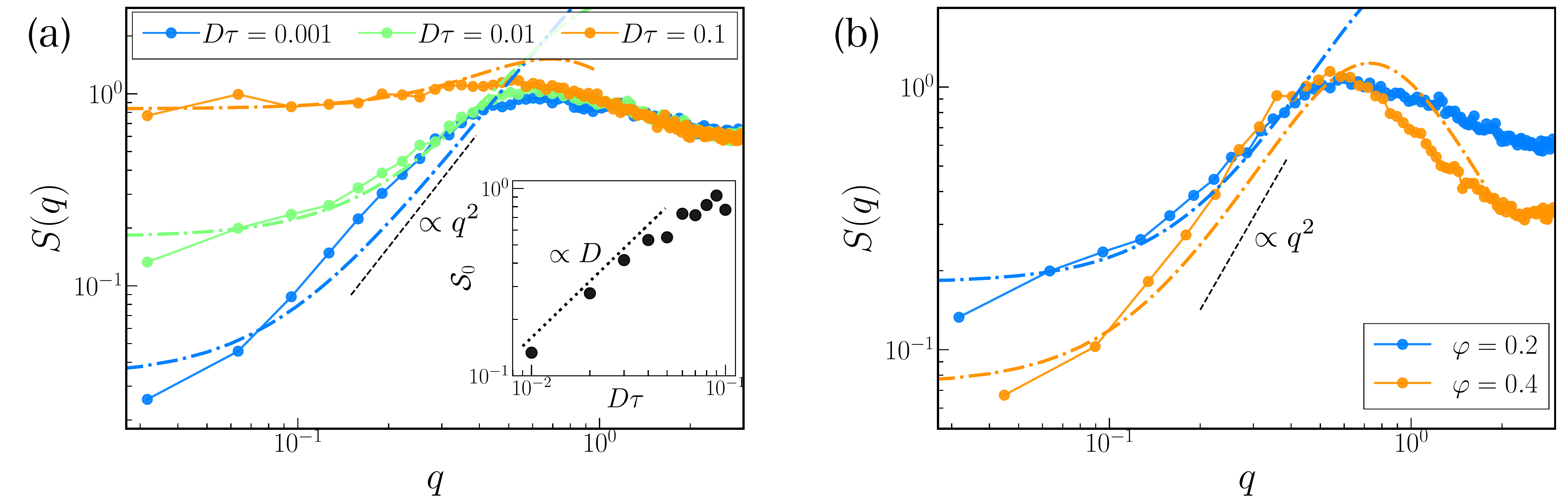}
 \caption{\label{fig4} 
 The static structure factor $S(q)$ for finite $D$. The filled circles indicate the numerical data and the dot-dashed lines are theoretical predictions by \eq{29}. The fitting parameter is only $b$.
 (a) $S(q)$ for various $D$ at $R=3\sigma$ and $\varphi=0.2$. 
  {The inset shows $\mathcal S_0 = S(q=0)$ as a function of $D\tau$.
 The dotted line is $\propto D$ as a guide for the eyes. }
 (b) $S(q)$ for $\varphi=0.2$ and $0.4$ at $R=3\sigma$ and $D\tau=0.01$.
 The black dashed lines stand for $\propto q^2$ as a guide for the eyes.
}
\end{figure*}
In Sec.~\ref{sim}, we compared the numerical simulation and our theoretical prediction only for $D=0$, at which exact hyperuniformity takes place. 
Here we show the numerical results and comparison with the theoretical prediction for finite $D$.
In this case, as shown in \eq{32}, our theory suggests that $S(q)$ remains a constant value in the limit $q\rightarrow 0$, {\it{i.e.}}, hyperuniformity does not appear. 
However, in case $\Omega = v_0/R > D$, $S(q)$ is suppressed for the low wavenumber regime. Hence, we focus only on the case of $\Omega > D$.
In Fig.~\ref{fig4}(a), we depict $S(q)$ for various $D$ at $R=3\sigma$ and $\varphi=0.2$. 
{The numerical data is represented by the filled circles.}
If $D$ is very small, $S(q)$ has the region where $S(q)\sim q^2$ even though $S(q\rightarrow 0)$ is constant. 
As $D$ increases, $S(q)$ has larger values in $q\rightarrow 0$, and the region where $S(q)\sim q^2$ disappears, meaning that the noise destroys hyperuniform structure.  
The dot-dashed line indicates the theoretical prediction by \eq{29}. 
$b$ is the sole fitting parameter.
The fitting range is $q<0.26$.
Unlike the case of $D\neq0$ shown in Sec.~\ref{sim}, the theoretical lines almost agree with the numerical data quantitatively. 
The inset of Fig.~\ref{fig4}(a) shows the $D$ dependence of $\mathcal S_0 = S(q=0)$. As $D$ decreases, $\mathcal S_0$ becomes smaller. The dotted line indicates $\mathcal S_0 \propto D$ predicted by the linearized theory, for small $D$ (see \eq{33}). 
Numerical data for small $D$ seem to obey this scaling.
We also show $S(q)$ for $\varphi=0.2$ and $0.4$ at fixed $D\tau=0.01$ and $R=3\sigma$ in Fig.~\ref{fig4}(b).
Interestingly, at $\varphi=0.4$, the theoretical prediction agrees with numerical data even for the peak at intermediate length scale.
Seeking the limitation of hydrodynamics and extending to more small-length scales are critical future works.

\newcommand{\newblock}{}
\bibliographystyle{apsrev4-2}
\end{document}